\documentclass[10pt,conference]{IEEEtran}
\usepackage{amsmath, amssymb, bm, cite, epsfig, psfrag}
\usepackage{graphicx}
\usepackage[top    = 0.7in,
bottom = 0.95in,
left   = 0.7in,
right  = 0.7in]{geometry}
\usepackage{dblfloatfix}
\usepackage{array}
\usepackage{textcomp}
\usepackage{longtable}
\usepackage{supertabular,booktabs}
\usepackage{multirow}
\usepackage[usenames,dvipsnames]{xcolor}

\usepackage{etoolbox}
\usepackage{pbox}
\usepackage{fixltx2e}
\usepackage{romannum}
\usepackage{filecontents}
\usepackage{enumerate}
\graphicspath{ {figures/} }
\usepackage{subfigure}
\usepackage{colortbl}
\usepackage{fancyhdr}

\usepackage{bm}
\newtoggle{conference}
\togglefalse{conference} %
\graphicspath{{figures/}}

\fancypagestyle{firststyle}{
	\fancyhf{}
	\fancyhead[L]{O. Kanhere and T. S. Rappaport, ``Millimeter Wave Position Location using Multipath Differentiation for 3GPP using Field Measurements,'' \textit{in GLOBECOM 2020 - 2020 IEEE Global Communications Conference,} Taipei, Taiwan, Dec. 2020, pp. 1--7.}     %

}
\usepackage{etoolbox}
\makeatletter
\patchcmd{\@makecaption}
{\scshape}
{}
{}
{}
\makeatletter
\patchcmd{\@makecaption}
{\\}
{.\ }
{}
{}
\makeatother

\begin{document}
\title{Millimeter Wave Position Location using Multipath Differentiation for 3GPP using Field Measurements}
\author{\IEEEauthorblockN{Ojas Kanhere, Theodore S. Rappaport\\}
	
\IEEEauthorblockA{	\small NYU WIRELESS\\
					NYU Tandon School of Engineering\\
					Brooklyn, NY 11201\\
					\{ojask, tsr\}@nyu.edu}}

\maketitle
\thispagestyle{firststyle}

\begin{abstract}

 3GPP air interface standards support meter-level position location of a user in a cellular network. With wider bandwidths and narrow antenna beamwidths available at mmWave frequencies, cellular networks now have the potential to provide sub-meter position location for each user. In this work, we provide an overview of 3GPP position location techniques that are designed for line-of-sight propagation. We discuss additional measurements required in the 3GPP standard that enable multipath-based non-line-of-sight position location. Further, we validate the concepts in this paper by using field data to test a map-based position location algorithm in an indoor office environment which has dimensions of 35 m by 65.5 m. We demonstrate how the fusion of angle of arrival and time of flight information in concert with a 3-D map of the office provides a mean accuracy of 5.72 cm at 28 GHz and 6.29 cm at 140 GHz, over 23 receiver distances ranging from 4.2 m to 32.3 m, using a single base station in line-of-sight and non-line-of-sight. We also conduct a theoretical analysis of the typical error experienced in the map-based position location algorithm and show that the complexity of the map-based algorithm is low enough to allow real-time implementation.

\end{abstract}
    

\section{Introduction}\label{Introduction}
 Position location (also called positioning or localization) is a key application for the fifth generation of mobile technologies (5G) and beyond \cite{Rappaport19a}. A variety of applications such as automated factories require precise knowledge of machinery and product locations \cite{Rap89c,Lu17} and will benefit from sub-meter position location. Self-driving cars of the future must be positioned with respect to their surroundings (i.e. other vehicles, pedestrians, and road structures). With accurate meter-level position location, commercial applications such as guided museum tours \cite{Kolodziej06a}, navigation in large malls or office spaces, locating products in retail stores \cite{Oplenskedal2019}, and patient tracking in hospitals are now realizable. 

Accurate position location of a user is critical for E911 emergency services \cite{Rap96a} and see-in-the-dark capabilities for firefighters and law enforcement. To ensure that the position of E911 callers is communicated to the first responders in an accurate and timely manner, the FCC has set forth regulatory positioning requirements of a horizontal positioning error $ < 50 $ m for 80\% of all wireless calls, and a vertical positioning error $ < 3 $ m in the top 25 US markets by April 3, 2021, with an end to end latency $ < 30 $ s \cite{FCC_2019}. A third generation partnership project (3GPP) study suggests a more stringent horizontal positioning requirement of $ < 3 $ m, with an end to end latency $ <1 $ s \cite{3GPP.38.855}, which may not be sufficient for a variety of commercial use cases that require sub-meter positioning accuracy. Advanced functionalities required by self-driving cars, such as overtaking, collision avoidance, and platooning require an accuracy of 30 cm \cite{Wymeersch2017}.

A large number of researchers are actively working towards achieving ubiquitous sub-meter position location accuracy. In \cite{Chen2017} the user equipment (UE) is localized by assuming that a line-of-sight (LOS) multipath component (MPC) was received by the UE, in addition to single-bounce non-line-of-sight (NLOS) MPCs. Simulations in a rectangular room with dimensions of 4 m by 3 m showed a localization error less than 20 cm when an antenna beamwidth of $ 8.6^\circ $ was used. Although work in \cite{Lin2018} did not assume the presence of a LOS path, the authors do assume that only single-bounce reflections reach the UE. The authors in \cite{Lin2018} derived an angular relationship between the angle of departure at the BS and the angle of arrival at the UE, by assuming that the environment is rectangular. Decimeter-level positioning error was achieved at an SNR of 0 dB at 28 GHz in simulations conducted in an indoor office and shopping mall with dimensions 15 $ \times $ 20 $ \times $ 4 m$ ^3 $ and 20 $ \times $ 50 $ \times $ 20 m$ ^3 $ respectively. The assumption that multi-bounce reflections are not feasible is not supported by measurements conducted at mmWave frequencies \cite{Mac15b,Kanhere19a}. 

The authors of \cite{Meissner13a} used ultra wideband (UWB) signals and UE tracking via an extended Kalman filter, without assuming single-bounce reflections. The TX position was mirrored across each obstruction in the environment to create ``virtual access points" (VAs) which act as additional anchor points to localize users. Four TXs were used to localize the user in an area of approximately 20 m $ \times $ 5 m.  Using a signal with pulse width $ T_p = 0.2$ ns, a root mean squared (rms) positioning error of 3.2 cm was achieved. 

The remainder of this paper is organized as follows. Section \ref{sec:3GPP} describes position location support in 3GPP. Map assisted positioning with angle and time (MAP-AT), a map-based positioning algorithm that fuses angular and temporal information with a map of the environment to provide centimeter-level position location in LOS and NLOS environments, is described in Section \ref{sec:MAP} and the complexity analysis and a theoretical error analysis is derived. The performance of MAP-AT on real-world indoor measurement data at 28 GHz and 140 GHz is evaluated in Section \ref{sec:results}. Concluding remarks and directions for future work are provided in Section \ref{sec:conclusion}.
\section{3GPP Position Location}\label{sec:3GPP}
Support for UE position location in the 3GPP standard can be traced back to the E-911 location service regulations set forth by the FCC in 1996 \cite{Rap96a,Rap98}. Today, with the ubiquitous nature of cell phones, 80\% E-911 calls are placed from wireless devices \cite{FCC_2019} and hence wireless position location is even more critical. Widespread adoption of a position location algorithm by network operators and cell phone manufacturers is only possible if the algorithm is standard-compliant. To develop standard-compliant position location algorithms, it is important to understand 3GPP position location, the wireless channel parameters currently used for 3GPP position location, as well as potential channel features that could be introduced in future 3GPP releases.

\begin{figure}
	\centering
	\includegraphics[width=0.5\textwidth]{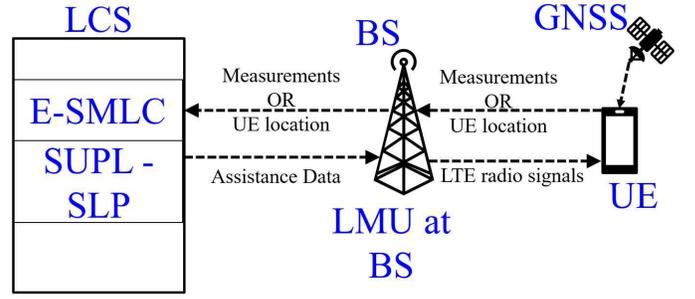}
	\caption{GNSS and LTE measurements are sent from the BS and UE to the LCS. If the UE location is calculated at the UE, the estimated position location solution is forwarded to the LCS.}
	\label{fig:architecture}

\end{figure}

 LTE UE position location is managed by a physical or logical entity called the location server (LCS), which obtains measurements from the BS and UE and provides assistance data (such as BS coordinates) to help position the UE. Two types of location solutions are offered by 3GPP LTE - control plane location solutions and secure user plane location (SUPL), each managed by a separate LCS. In control plane positioning, managed by the Evolved Serving Mobile Location Center (E-SMLC), the network routes positioning messages over signaling connections to ensure quick and secure positioning. In SUPL, managed by the SUPL secure location platform (SUPL-SLP), message exchange takes place over the data traffic link. The location measurement unit (LMU) at the BS measures the sounding reference signal (SRS) required for uplink positioning, while the positioning reference signal (PRS) is measured at the UE for downlink positioning. The overall LTE positioning architecture is illustrated in Fig. \ref{fig:architecture}.

\subsection{Position Location Techniques Supported by 3GPP}

\begin{figure}
	\centering
	\includegraphics[width=0.4\textwidth]{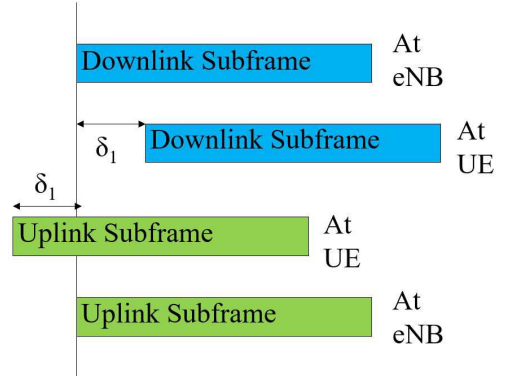}
	\caption{TA ($ \delta_1 $) ensures uplink frames of all UEs are aligned.}
	\label{fig:timing_advance}
\end{figure}

The simplest 3GPP position location technique, Cell ID (CID) proximity positioning, estimates the UE to be at the serving BS. With Enhanced CID (E-CID), CID accuracy is improved by incorporating reference signal strength (RSS) measurements at the UE, and timing advance (TA) and AoA measurements at the BS \cite{3GPP.25.305}. Due to the varying distances of UEs from the BS, the uplink data arrives at the BS with a delay proportionate to the BS-UE distance. To ensure uplink frame time-alignment, the TA is sent as downlink feedback and enables UEs to adjust their uplink transmission, and thus provides an estimate of the round trip time (RTT) of the first arriving MPC at the BS, as seen in Fig. \ref{fig:timing_advance}. The time of flight (ToF) of the first arriving MPC (the one-way travel time) is equal to half the RTT. The minimum reportable one-way distance $ d_{min} $, calculated from the ToF, decreases with an increase in subcarrier spacing (SCS) and is given by \cite{3GPP.38.213}
\begin{align}
d_{min} = 78.12/2^\mu,
\end{align}
when the SCS is $ 2^\mu \times 15$ kHz. In 5G-NR, the maximum SCS is 60 kHz for lower frequency bands (below 6 GHz), corresponding to a minimum reportable distance of 19.52 m, while the maximum SCS is 480 kHz for higher frequency bands (above 24 GHz), corresponding to a minimum reportable distance of 2.44 m.

Although specific details of how to utilize the RSS and TA measurements are not provided in the standard, TA and/or RSS may be utilized to estimate the distance between the BS and UE \cite{Kanhere18a, Kanhere19a}. The distance estimate may be combined with AoA to calculate the position of the UE via simple geometric calculations \cite{Kanhere18a, Kanhere19a}.

While antenna arrays at BS are becoming more prevalent due to the need for beamforming at mmWave frequencies, some current BSs may not be fitted with antenna arrays due to cost considerations. In the absence of AoA, multiple distance estimates (estimated from RSS/TA) may also be used for UE position location (via trilateration) if the UE is in the coverage area of at least three BSs. The primary BS may initiate forced handovers, allowing the UE to estimate the TA/RSS from other neighboring BSs \cite{Rap98}.

GNSS receivers (RXs), present in nearly all modern cellular devices, localize a UE to within 5 m when four or more satellites are directly visible\cite{Diggelen_2015}. However, in urban canyons where the direct path to satellites is blocked, GNSS performance deteriorates. With assisted GNSS (A-GNSS), 3GPP networks improve GNSS performance by providing assistance information to the UE that helps improve RX sensitivity and reduce time to first fix (TTFF) and UE power consumption \cite{3GPP.36.305}. The cellular network provides external information that improves the GNNS position TTFF by utilizing a coarse estimate of the UE location (for instance via E-CID) to reduce the frequency/code-delay search space. The lower TTFF allows UEs to consume less power as the GNSS RX need not be always-on.

Downlink TDoA (called observed time difference of arrival (O-TDOA) in 3GPP \cite{3GPP.25.305}) is measured at UE. The difference in time at which the PRS is received at the UE from two BSs is called the reference signal timing difference (RSTD). Since the BSs are synchronized via GNSS satellites, RSTD has a direct relation to the geometric difference in distance of the two BSs from the UE. O-TDoA supports a time resolution of  0.5 $T_s $ (4.88 m), when RSTD $ \leq$ 4096 $ T_s $ and a time resolution of 1 $ T_s $ (9.67 m) when 4096 $ T_s $ $ \leq $ RSTD $ \leq $ 15391 $ T_s $, where 1 $ T_s $ = 32.522 ns (9.76 m). 

Just as O-TDoA is measured at the UE, uplink time difference of arrival (UTDOA) is measured at the BS, allowing UEs lacking capabilities to make OTDoA measurements to be localized. The SRS, a Zadoff-Chu sequence transmitted by the UE, is utilized by two or more pairs of BS to measure the relative time of arrival. A minimum resolution of 2 $ T_s $ (19.51 m) is possible.

To improve the vertical component of UE position location, the barometric positioning method is used, wherein the atmospheric pressure at the UE is measured using barometric pressure sensors found in most modern cell phones \cite{3GPP.25.305}. Since the atmospheric pressure decreases with an increase in UE height, by calibrating the barometric pressure sensor to the atmospheric pressure at a known height, the vertical UE position may be determined.
\subsection{Support for Position Location algorithms of the future}
With the advent of ultra-wide bandwidths due to the utilization of mmWave frequency bands, MPCs can now be resolved to a finer time resolution. Localization techniques that exploit multipath information require delay and angle measurements of more than one path. Currently, 3GPP supports the reporting of the relative delay of all MPCs with respect to the MPC utilized to calculate the RSTD, via the additional path information element with a resolution of 0.5 $ T_s $ (4.88 m)\cite{3GPP.36.133}. By adding the relative MPC delay to TA, the absolute time of arrival of individual MPCs can be calculated. 3GPP currently supports the measurement of the AoA of only one signal at the BS, due to which AoA information of individual MPCs is lacking.

A consequence of moving to higher frequencies is the development of electrically large (yet physically small) antenna arrays at the BS and UE. It is expected that future releases will include AoA measurements at the UE since multiple input multiple output (MIMO) capable UEs are already widely available. AoA measurements at the UE would also require orientation estimation which could be done via in-built sensors present in the UE, such as the accelerometer, magnetometer, and gyroscope \cite{mohssen2014}. 
\section{Position Location with MAP-AT}\label{sec:MAP}
MAP-AT is a position location algorithm that utilizes a map of the environment (pre-generated or generated on-the-fly) along with temporal and angular information. The angle of arrival (AoA) and time of flight (ToF) of multipath components are estimated at the BS. As discussed in Section \ref{sec:3GPP}, TA, conventionally used to synchronize the uplink transmissions of UEs, is used to estimate the ToF of the first arriving MPC at the BS. Additional path information, measured on the downlink at the UE, provides the relative ToF of later arriving MPCs with respect to the first arriving MPC. By adding the TA measured at the BS to the relative path delays, the absolute delays of the MPCs arriving at the UE can be estimated. By channel reciprocity, the absolute delays of each MPC arriving at the BS on the uplink are equal to the absolute delays of the MPCs arriving at the UE on the downlink. Thus a provision to estimate the ToF of MPCs, as required by MAP-AT, is already supported by 3GPP. However, the AoA of the MPCs arriving at the BS must also be measured, which is currently not supported by 3GPP (current support is only for the measurement of a single AoA at the BS).

Since mmWave signals may suffer multiple reflections \cite{Ju20}, MAP-AT makes no assumptions on the number of reflections or transmissions of an MPC before the signal reaches the UE. After one reflection/transmission, if the ToF and AoA of a multipath signal sent from the BS are known, there are two possible locations of the UE. If the signal reached the user after one reflection, the UE and BS must lie on the same side of the reflecting object. If the signal reached the BS directly from the UE, or through one obstruction, the BS and UE must lie on opposite sides of the obstruction. The possible locations of the user, based on ToF and AoA at the BS shall henceforth be referred to as \textit{candidate locations}.

\begin{figure}
	\centering
	\includegraphics[width=0.35\textwidth]{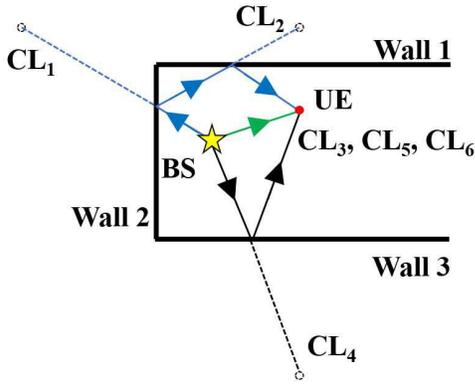}
	\caption{ Three MPCs arrive at the UE shown above - one LOS component (in green) and two NLOS components (in blue and black). Of the six candidate locations for the user, based on AoA and ToF measurements at the BS ($ \text{CL}_1 - \text{ CL}_6 $), three candidate locations $ (\text{CL}_3,\text{ CL}_5, \text{ CL}_6) $ correspond to the actual location of the user\cite{Kanhere19a}. }
	\label{fig:candidates}
\end{figure}

When the signal is reflected or transmitted multiple times, each successive reflection/transmission creates more candidate locations, as seen in Fig. \ref{fig:candidates}. The process of finding candidate locations is repeated for all MPCs. If a UE in LOS environment receives a single MPC, since the MPC corresponds to a single candidate location, the UE location can be unambiguously determined as the position of the candidate location. In NLOS, since a single MPC will create two or more candidate locations, the BS is not able to determine which candidate location corresponds to the user's true location. However, when two or more MPCs arrive at the user, a majority of the candidate locations will correspond to the true location of the user. For each MPC arriving at the user, one candidate location calculated based on the AoA and ToF of the MPC corresponds to the true user's location. Fig. \ref{fig:candidates} depicts all the candidate locations when three MPCs are received by the UE from the BS. The location of the UE corresponds to the candidate location identified by the maximum number of MPC.

\subsection{Theoretical Error Analysis}\label{sec:error}
\begin{figure}
	\centering
	\includegraphics[width=0.35\textwidth]{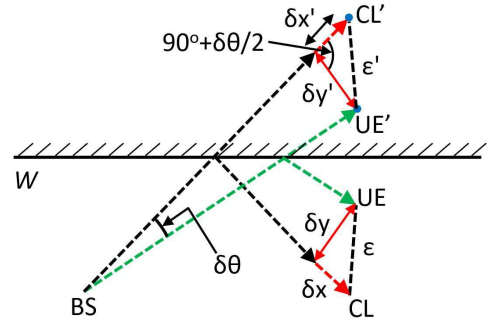}
	\caption{ When the angular error $ \delta \theta $ is small, the angular and temporal components of position location error ($ \delta y $ and $ \delta x $ respectively) of an MPC are near-orthogonal, making an angle of $90^\circ + \delta\theta/2  $. The magnitude of $ \varepsilon $ and $ \varepsilon' $ are equal due to which error analysis may be carried out by assuming all MPCs are LOS.}
	\label{fig:perpendicular_errors}
\end{figure}

Before considering the overall position location error of MAP-AT, we shall first consider the position estimation error of a single candidate location. Let $ \varepsilon $ be the vector between the true UE location and the candidate location. Consider an MPC with ToF $ t $ which is measured by the BS with an angular error of $ \delta \theta $ and a temporal error of $ \delta t $. Appealing to the physical principle that images of objects in plane mirrors have the same length as the objects themselves, it is clear that $ \varepsilon $ (the length of the error vector) and $ \varepsilon' $ (the length of the image obtained after reflection through a wall W) have the same magnitude, as seen in Fig. \ref{fig:perpendicular_errors}. Indeed, even after multiple reflections, the length of the error vector remains constant. As a result of this observation, error analysis of any NLOS MPC with ToF $ t $ may now be reduced to the error analysis of a LOS MPC with ToF $ t $, angular error of $ \delta \theta $ and a temporal error of $ \delta t $. The distance covered by the MPC is given by $ r = c\times t $, where $ c $ is the speed of light.

As seen in Fig. \ref{fig:perpendicular_errors}, for small angular errors angular and distance errors combine in a near-orthogonal manner, due to which, if the angular component of position location error is given by $\delta x =  r \times \delta \theta $ and the temporal component of position location error is given by $\delta y= c \times \delta t $, the magnitude of the error vector is given by $ \varepsilon = \sqrt{\delta x^2+ \delta y^2} $. Assuming the angular and temporal estimation errors $ \delta x $ and $ \delta y $ are Gaussian, $ \varepsilon $ is a generalized chi-distribution, with a second moment equal to the sum of the variances of the angular and temporal estimation errors. Note that if the variances of $ \delta x $ and $ \delta y $ were the same, $ \varepsilon $ would be a Rayleigh random variable. However in general $ \varepsilon $ is a chi-distribution. The mean position location error was estimated from Monte Carlo simulations, and is equal to 7.56, 10.18, and 16.26 cm when $ r =  $ 5, 10, and 20 m. Temporal and angular noise were zero-mean Gaussian random variables with standard deviations of 0.25 ns and $ 0.5^\circ $ respectively.

In MAP-AT, an estimate of the UE location is found by calculating the centroid of candidate locations belonging to the largest cluster. Thus, assuming that there are N candidate locations in the largest cluster and that each of the MPC error vectors are \textit{i.i.d} with mean $ \mu $ and variance $ \sigma^2 $, the position location error magnitude is also Chi distributed with mean $ \mu $ and variance $ \sigma^2/N $.
\subsection{Computational Complexity}

As discussed above, the estimation of AoA and ToF in MAP-AT does not require additional computations at the BS or UE, beyond the requirements for wireless communication. Once the AoA and ToF are estimated, the candidate locations must be calculated. Unlike ray tracing, where signal paths in all directions must be calculated, MAP-AT requires path calculations only in the directions where the MPCs arrive from. Indoor mmWave channel statistics at 28 an 140 GHz provided in \cite{Ju20} indicate a maximum of 10 MPCs, with a median of fewer than four MPCs per location. Typically, mmWave signals suffer fewer than 3 reflections. In general, with a maximum of $ k $ reflections and $ M $ MPCs, $ 2^k \cdot M $ reflections must be calculated, which can be done in real-time on any modern UE or BS.

To further reduce computations, the temporal and angular domain may be quantized. The candidate locations could then be pre-computed and stored as a look-up table, depending on the AoA and ToF of the MPC.

Close-by candidate locations are grouped together. Let there be $ n $ distinct candidate locations. Grouping all $ n $ candidate locations will involve at most ${n\choose 2}$ euclidean distance calculations, i.e. a computational complexity of $ \mathcal{O}( n^2). $

In simulations conducted on an Intel i7-3770 CPU with 16 GB RAM, computing all the candidate locations for one MPC took 13.01 $ \mu  $s on average while grouping candidate locations required an additional 1.29 $ \mu $s. Since candidate locations for all MPCs may be computed in parallel, the total computation time of MAP-AT was 14.30 $ \mu $s per location. with sequential calculation of candidate locations, assuming four MPCs per location, MAP-AT only requires 53.33 $ \mu $s per location, which is low enough to enable real-time implementation.

\section{MAP-AT performance with Real-world Indoor mmWave measurements}\label{sec:results}
\begin{figure*}
	\centering
	\includegraphics[width=0.8\textwidth]{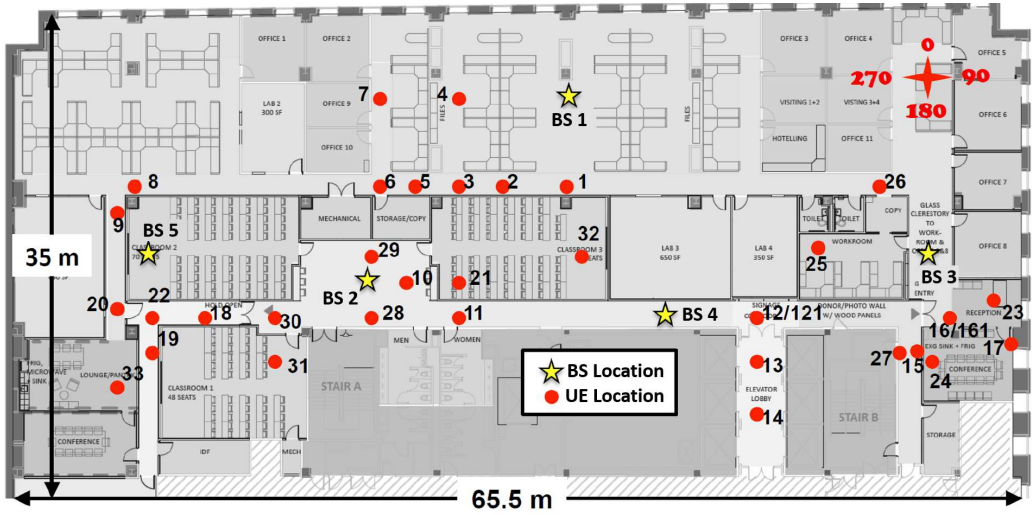}
	\caption{Map of the 9th floor of 2 MTC, depicting the indoor locations where measurements were conducted at 28 GHz and 140 GHz \cite{Mac15b,Ju20}}
	\label{fig:indoor_locations}
\end{figure*}

The localization performance of MAP-AT shall now be examined with real-world indoor measurements at 28 and 140 GHz. Measurements at both frequencies were conducted in the same indoor environment, on the $ 9^{th} $ floor of 2 MetroTech Center, the former research center of NYU WIRELESS. The floor plan is provided in Fig. \ref{fig:indoor_locations}. A wideband sliding correlator-based channel sounder was used in both measurement campaigns, having a null-to-null RF bandwidth of 800 MHz and 1 GHz at 28 GHz and 140 GHz respectively.

\begin{figure}
	\centering
	\includegraphics[width=0.4\textwidth]{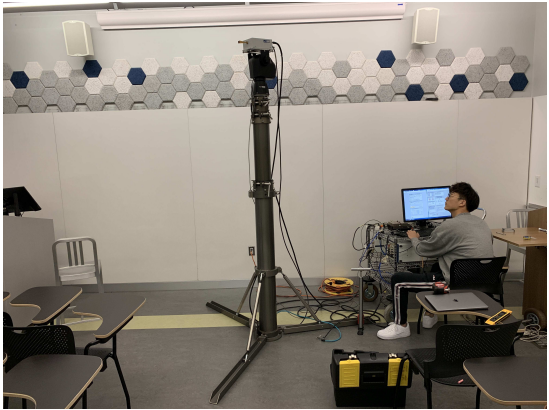}
	\caption{BS 5 was mounted on a gimbal at a height of 2.5 m in a classroom.}
	\label{fig:TX_5}
\end{figure}

\begin{figure}
	\centering
	\includegraphics[width=0.4\textwidth]{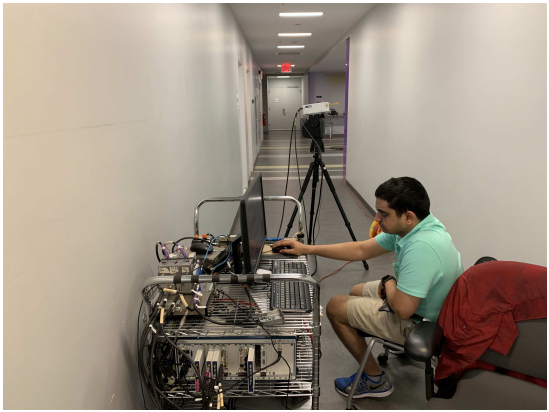}
	\caption{UE 11 was mounted on a gimbal at a height of 1.5 m in a long corridor.}
	\label{fig:RX_11}
\end{figure}

Identical horn antennas at the transmitter (TX) and RX with 15 dBi (27 dBi) gain and $ 30^\circ (8^\circ)$ half-power beamwidths (HPBW) were used at 28 GHz (140 GHz). The horn antennas were mounted on electronically steerable gimbals with sub-degree accuracy in the azimuth and elevation plane and rotated in steps of the antenna HPBW. The TX gimbal was at a height of 2.5 m, just below the ceiling (2.75 m high) to replicate the location where mmWave BS could be deployed, as seen in Fig. \ref{fig:TX_5} while the RX gimbal was at a height of 1.5 m, the typical mobile UE height as seen in Fig. \ref{fig:RX_11}.
To detect MPCs as required by MAP-AT, after the TX and RX antennas were pointed directly at each other (boresight), the RX antenna was swept in the azimuth plane in HPBW steps. To measure additional MPCs, the TX and RX antennas were realigned at boresight and the TX antenna was swept in the azimuth plane in HPBW steps. 

Spatial lobes in the TX and RX antenna sweeps were extracted by defining a threshold 10 dB below the maximum power. All contiguous angular directions with powers above the 10 dB threshold were considered to belong to the same spatial lobe \cite{Samimi2015}. The mean angle of each spatial lobe ($ \mu_\theta $) was calculated as a weighted average of all the segments belonging to the spatial lobe \cite{3GPP.38.901}, given by:
 \begin{align}
\mu_\theta = \arg\left( \sum_{i} P_i\exp(j\theta_i) \right)  ,
 \end{align}
 where $ \theta_i $ and $ P_i $ are the AoA and power of MPC $ i $ respectively. 

Due to the lack of ToF measurements, the measured AoA was augmented with ToF predicted by NYURay, a 3-D mmWave ray tracer \cite{Kanhere19a}. NYURay is a hybrid ray tracer calibrated to real-world mmWave measurements capable of providing accurate temporal, angular, and power measurements. Since an exhaustive search of all TX-RX pointing angles was impractical for the real-world measurements, some MPCs predicted by NYURay were not detected in the measurements (a median of fewer than 4 MPCs were measured at each location at 28 and 140 GHz \cite{Ju20}). Zero mean Gaussian noise with a standard deviation of 0.25 ns and $ 0.5 ^\circ $ was added to the measured ToF and AoA respectively to model measurement uncertainty. In order to ensure that meaningful statistics were obtained, 1000 simulation runs were conducted for each BS-UE pair. 

 Good localization results were obtained using a single BS at each frequency. A mean error of 5.72 cm was observed over 23 locations at 28 GHz, while a mean error of 32.5 cm was observed over 10 locations at 140 GHz. Although the performance of MAP-AT at both frequencies was good, the relatively worse performance at 140 GHz was caused by a single outlier at 140 GHz, located behind a corner where only 2 MPCs were detected, due to which a small error in AoA estimation at the BS led to an incorrect prediction of the reflecting obstruction resulting in an outlier error of 2.68 m which skewed the mean error at 140 GHz. However, 3 MPCs were detected at 28 GHz, and since MAP-AT relies on majority candidate location selection, the UE location was correctly predicted. A mean error of 6.29 cm was obtained over the remaining 9 locations at 140 GHz. The theoretical analysis from Section \ref{sec:error} predicted a mean position location error of 10.18 cm at 10 m, which is in good agreement with the mean errors obtained from measurements. Table \ref{table:28GHz} and \ref{table:140GHz} illustrate how the error varies with frequency and BS-UE separation distance in LOS and NLOS environments at 28 GHz and 140 GHz respectively.

\begin{table}[]
	\centering
	\caption{Performance of the MAP-AT at 28 GHz for different BS-UE separation distances in LOS and NLOS environments. }\label{table:28GHz}
\setlength\tabcolsep{4pt}
	\begin{tabular}{|c|c|c|c|c|c|c|}
	\hline
	\begin{tabular}[c]{@{}c@{}}\\	BS-UE  \\    distance \end{tabular}  & \begin{tabular}[c]{@{}c@{}}\\	Number of   \\    UE Locations  \end{tabular}                                                &  \begin{tabular}[c]{@{}c@{}}\\$ \mu_d $ \\    (m)\end{tabular}       &  \begin{tabular}[c]{@{}c@{}}\\$ \sigma_d $ \\    (m)\end{tabular}      & \begin{tabular}[c]{@{}c@{}}\\	BS-UE   \\   Link type \end{tabular}                       & \begin{tabular}[c]{@{}c@{}}\\$ \mu_\varepsilon $ \\    (cm)\end{tabular} & \begin{tabular}[c]{@{}c@{}}\\$ \sigma_\varepsilon $ \\    (cm)\end{tabular}\\ \hline
	\textless 10 m & \begin{tabular}[c]{@{}c@{}}5\\   8\end{tabular} & \begin{tabular}[c]{@{}c@{}}6.10\\   8.02\end{tabular} & \begin{tabular}[c]{@{}c@{}}1.41\\   1.07\end{tabular}   &\begin{tabular}[c]{@{}c@{}}LOS\\   NLOS\end{tabular}&\begin{tabular}[c]{@{}c@{}}5.95\\   5.36\end{tabular} &   \begin{tabular}[c]{@{}c@{}}1.46\\   1.00\end{tabular}                \\ \hline
	 10 - 35 m & \begin{tabular}[c]{@{}c@{}}3\\   7\end{tabular} & \begin{tabular}[c]{@{}c@{}}15.71\\   16.10\end{tabular} & \begin{tabular}[c]{@{}c@{}}4.87\\   7.37\end{tabular}   &\begin{tabular}[c]{@{}c@{}}LOS\\   NLOS\end{tabular}&\begin{tabular}[c]{@{}c@{}}6.40\\   5.67\end{tabular} &   \begin{tabular}[c]{@{}c@{}}1.96\\   1.19\end{tabular}                \\ \hline
	(all) & \begin{tabular}[c]{@{}c@{}}8\\   15\end{tabular} & \begin{tabular}[c]{@{}c@{}}9.70\\   11.79\end{tabular} & \begin{tabular}[c]{@{}c@{}}5.71\\   6.43\end{tabular}   &\begin{tabular}[c]{@{}c@{}}LOS\\   NLOS\end{tabular}&\begin{tabular}[c]{@{}c@{}}6.12\\   5.51\end{tabular} &   \begin{tabular}[c]{@{}c@{}}1.54\\   1.07\end{tabular}                \\ \hline
\end{tabular}
\end{table}

\begin{table}[]
	\centering
	\caption{Performance of the MAP-AT at 140 GHz for different BS-UE separation distances in LOS and NLOS environments. }\label{table:140GHz}
	\setlength\tabcolsep{4pt}
	\begin{tabular}{|c|c|c|c|c|c|c|}
		\hline
		\begin{tabular}[c]{@{}c@{}}\\	TX-RX  \\    distance \end{tabular}  & \begin{tabular}[c]{@{}c@{}}\\	Number of   \\    UE Locations  \end{tabular}                                                &  \begin{tabular}[c]{@{}c@{}}\\$ \mu_d $ \\    (m)\end{tabular}       &  \begin{tabular}[c]{@{}c@{}}\\$ \sigma_d $ \\    (m)\end{tabular}      & \begin{tabular}[c]{@{}c@{}}\\	BS-UE   \\   Link type \end{tabular}                       & \begin{tabular}[c]{@{}c@{}}\\$ \mu_\varepsilon $ \\    (cm)\end{tabular} & \begin{tabular}[c]{@{}c@{}}\\$ \sigma_\varepsilon $ \\    (cm)\end{tabular}\\ \hline
		\textless 10 m & \begin{tabular}[c]{@{}c@{}}3\\   1\end{tabular} & \begin{tabular}[c]{@{}c@{}}6.32\\   8.34\end{tabular} & \begin{tabular}[c]{@{}c@{}}2.38\\   -\end{tabular}   &\begin{tabular}[c]{@{}c@{}}LOS\\   NLOS\end{tabular}&\begin{tabular}[c]{@{}c@{}}6.86\\   5.21\end{tabular} &   \begin{tabular}[c]{@{}c@{}}1.38\\   -\end{tabular}                \\ \hline
	 10 - 30 m & \begin{tabular}[c]{@{}c@{}}1\\   5\end{tabular} & \begin{tabular}[c]{@{}c@{}}13.06\\   15.14\end{tabular} & \begin{tabular}[c]{@{}c@{}}-\\   7.12\end{tabular}   &\begin{tabular}[c]{@{}c@{}}LOS\\   NLOS\end{tabular}&\begin{tabular}[c]{@{}c@{}}7.56\\   58.35\end{tabular} &   \begin{tabular}[c]{@{}c@{}}-\\   117.46\end{tabular}                \\ \hline
(all) & \begin{tabular}[c]{@{}c@{}}4\\   6\end{tabular} & \begin{tabular}[c]{@{}c@{}}8.00\\   14.00\end{tabular} & \begin{tabular}[c]{@{}c@{}}3.89\\   6.94\end{tabular}   &\begin{tabular}[c]{@{}c@{}}LOS\\   NLOS\end{tabular}&\begin{tabular}[c]{@{}c@{}}7.04\\   49.50\end{tabular} &   \begin{tabular}[c]{@{}c@{}}1.18\\   107.28\end{tabular}                \\ \hline
	\end{tabular}
\end{table}

\section{Conclusion}\label{sec:conclusion}
This paper has provided an overview of the position location techniques supported by 3GPP. By adding AoA information in the multipath report in addition to the relative multipath delays, novel multipath-based position location schemes could be made compatible with the standard. The computational complexity of MAP-AT, a map-based position location algorithm that fuses AoA and ToF information, is examined and it is shown that MAP-AT is a light-weight algorithm that can be implemented in real-time. Theoretical error analysis shows that the position error is a generalized Chi distributed random variable. The performance of MAP-AT is evaluated with real-world mmWave data at 28 GHz and 140 GHz. A mean position location error of 5.72 cm was obtained at 28 GHz over 23 UE locations, while a mean error of 6.29 cm was obtained at 140 GHz over 9 UE locations, after removing one outlier. Future work shall focus on vehicular and outdoor UE position location. 

\section{Acknowledgments}
This material is based upon work supported by the NYU WIRELESS Industrial Affiliates Program and National Science Foundation (NSF) Grants: 1702967 and 1731290.
\bibliographystyle{IEEEtran}
\bibliography{main}{}

\end{document}